\documentclass[12pt]{article}
\usepackage{graphicx}
\usepackage{epsf,amsmath,bbold,amsfonts,stmaryrd}

\usepackage[utf8]{inputenc}
\usepackage{caption}
\usepackage{mathrsfs}
\usepackage{appendix}
\usepackage{amssymb}
\usepackage{float}
\usepackage{color} 
\usepackage{cite}
\usepackage[colorlinks]{hyperref}
\hypersetup{pageanchor=false}
\usepackage{indentfirst}
\usepackage{url}
\usepackage{float}
\usepackage{caption}
\usepackage[numbers,square,comma,sort&compress,merge]{natbib} 
\usepackage{comment}
\usepackage{mathtools}

\usepackage{footnote}
\makesavenoteenv{tabular}
\makesavenoteenv{table}

\usepackage{booktabs}
\usepackage{siunitx}

\def\a{\alpha}
\def\b{\beta}

\renewcommand{\a}{\alpha}
\renewcommand{\b}{\beta}

\renewcommand{\l}{\lambda}

\newcommand{\jh}{\xi_h}
\newcommand{\jx}{\xi_\chi}

\newcommand{\x}{\chi}

\newcommand{\be}[1]{\begin{equation} \centering \label{#1}}
\newcommand{\ee}{\end{equation}}

\def\bea{\begin{eqnarray}}
\def\eea{\end{eqnarray}}

\def\ba{\begin{array}}
\def\ea{\end{array}}

\def\bc{\begin{center}}
\def\ec{\end{center}}

\def\bl{\begin{flushleft}}
\def\el{\end{flushleft}}

\def\br{\begin{flushright}}
\def\er{\end{flushright}}

\def\bi{\begin{itemize}}
\def\ei{\end{itemize}}

\def\bt{\begin{tabular}}
\def\et{\end{tabular}}

\newtheorem{question}{Question}
\def\bq{\begin{question}}
\def\eq{\end{question}}

\newtheorem{definition}{Def}
\def\bd{\begin{definition}}
\def\ed{\end{definition}}

\newtheorem{answer}{Answer}
\def\ban{\begin{answer}}
\def\ean{\end{answer}}

\newtheorem{possibleanswer}{Possible answer}
\def\bpa{\begin{possibleanswer}\normalfont}
\def\epa{\end{possibleanswer}}

\newtheorem{theorem}{Theorem}
\def\bth{\begin{theorem}}
\def\eth{\end{theorem}}

\begin{document}

\begin{titlepage}
\vspace{5cm}

\vspace{2cm}

\begin{center}
\bf \Large{On the geometrical interpretation of scale-invariant models of inflation}

\end{center}

\begin{center}
{\textsc {Georgios K. Karananas$\,^\dagger$, Javier Rubio$\,^\ddagger$}}
\end{center}

\begin{center}
{\it $^\dagger$Laboratory of Particle Physics and Cosmology, \\
Institute of Physics, \\
\'Ecole Polytechnique F\'ed\'erale de Lausanne, \\ 
CH-1015, Lausanne, Switzerland\\
\leavevmode\\
$^\ddagger$Institut f\"ur Theoretische Physik,\\
Universit\"at Heidelberg,\\
Philosophenweg 16, D-69120 Heidelberg}
\end{center}

\begin{center}
\texttt{\small georgios.karananas@epfl.ch} \\
\texttt{\small j.rubio@thphys.uni-heidelberg.de} 
\end{center}

\vspace{2cm}

\begin{abstract}

We study the geometrical properties of scale-invariant two-field models of inflation. 
In particular, we show that when the field-derivative space in the Einstein frame is maximally symmetric 
during inflation, the inflationary predictions can be universal and independent of the details of the theory.

\end{abstract}

\end{titlepage}

\section{Introduction}

The accurate measurements of the cosmic microwave background~\cite{Ade:2015lrj} have established
inflation as the leading paradigm for explaining the background properties of the observable Universe and
the origin of the primordial perturbations giving rise to structure formation 
\cite{Starobinsky:1980te,Mukhanov:1981xt,Guth:1980zm,Linde:1981mu,Albrecht:1982wi,Linde:1983gd}.

The conditions for inflation are usually formulated as conditions on the flatness 
of the potential of a canonically normalized scalar field. Note however that noncanonical kinetic terms 
are ubiquitous in nonminimally coupled theories of inflation  
\cite{Adler:1982ri,Minkowski:1977aj,Smolin:1979uz,Zee:1978wi,Salopek:1988qh,Bezrukov:2007ep} when these theories are formulated 
in the Einstein frame. For models involving a single field, the complexity in the noncanonical kinetic 
term can be easily reabsorbed in the form of the potential by performing a field redefinition. The situation changes 
completely if more than a scalar field is nonminimally coupled to gravity. When this happens, the predictions 
of the model are generically affected by the Einstein-frame kinetic mixing among the fields, even if the 
inflationary potential is dominated by a single component.

An interesting subset of nonminimally coupled inflationary models are those displaying global scale invariance, i.e. 
invariance under the transformations
\be{scalinv}
x^\mu\rightarrow \a^{-1}x^\mu\ ,\hspace{10mm}\Phi_i(x)\rightarrow \a^{d_i}\Phi_i(\a^{-1} x) \ ,
\ee
with $\a$ a constant, $\Phi_i$ the fields of the theory and $d_i$ their corresponding mass dimension. The presence of such a symmetry can be quite appealing, since all scales at the classical level are generated dynamically and can be sourced by the spontaneous breaking of dilatations~\cite{Shaposhnikov:2008xb,Shaposhnikov:2008xi}. This common 
origin of the various dimensionful parameters might give us some insight, and eventually even an answer, 
to the long-standing question regarding the smallness of the Higgs mass and the cosmological constant as compared 
to the Planck mass $M_P=2.4\times 10^{18}$~GeV~\cite{Bardeen:1995kv,Meissner:2006zh,Shaposhnikov:2008xb,Shaposhnikov:2008xi}. As for model-building, scale invariance is also a powerful tool, since 
the Lagrangian describing the dynamics of the theory under consideration is subject to the selection rules imposed by symmetry.

The simplest model within the scale-invariant  category is the induced gravity scenario~\cite{Zee:1978wi}. In spite of its simplicity, this 
model is excluded by observations since it does not allow for a graceful inflationary exit. In order to construct 
viable scale-invariant theories of inflation, it seems unavoidable to introduce at least two scalar degrees of freedom, one of
which should be thought of as a dilaton. This additional \textit{dynamical} field can either be introduced \textit{ad hoc}, or emerge 
naturally from some physical requirement.

In this work, we will consider two-field models of inflation which are invariant under~\eqref{scalinv} and with kinetic terms that 
are at most quadratic in derivatives. It turns out that the most general theory satisfying these conditions, involves
a number of \textit{a priori} independent functions, which, for dimensional reasons, depend only on one of the fields. 
Making general statements without specifying the exact form of these theory defining functions is certainly not feasible. 
Nevertheless, it might after all be possible to overcome this obstacle, provided that there exist some constraints
that enable us to relate them in a nontrivial manner. As we will show, the geometry of the two-dimensional target manifold associated with the kinetic part of
the theory plays a central role here. If for the field values relevant for inflation its curvature is 
approximately constant, then the field-derivative space is maximally symmetric. This translates into a
differential equation that can be used to express the whole kinetic sector in terms of the function that appears in front of  the dilaton's 
kinetic term and its derivatives.

The paper is organized as follows. In Sec.~\ref{sec:IG}, we consider a 
two-field scale-invariant model of inflation in which only one of the two fields displays nontrivial interactions. 
After discussing the limitations of this induced gravity scenario for obtaining a graceful inflationary exit, we 
consider a minimal extension of the model containing nontrivial interactions for both scalar fields. A 
detailed analysis of this theory with special emphasis on its 
geometrical structure appears in Sec.~\ref{sec:HDmodel}. The isolation of the main elements contributing  
to the inflationary observables in this minimal extension will allow us to generalize the results to
a broad class of scale-invariant theories. This is done in Sec.~\ref{sec:gencase}. We present 
our conclusions in Sec.~\ref{sec:concl}.

\section{Induced gravity}\label{sec:IG}
 As a warm up, we start with a two-field scale-invariant model in which one of the fields is interacting, whereas
the other has only a kinetic term. In particular, let us consider an induced gravity model whose dynamics is described by the following Lagrangian density\,\footnote{In order to shorten
the expressions we suppress the Lorentz indices. The implicit contractions should be understood in terms 
of the metric associated with the frame under consideration.} 
\be{Hig-Inf-Ind2}
\frac{\mathscr L}{\sqrt{g}}=\frac{f(h)}{2}R-\frac{1}{2}(\partial h)^2-\frac{1}{2}(\partial \chi)^2-
U(h) \ ,
\ee
with $g=-\det(g_{\mu\nu})$, and
\be{}
f(h)=\xi_h h^2\ ,\hspace{10mm}U(h)=\frac{\lambda}{4} \,h^4\ .
\ee
The non-minimal coupling $\xi_h$ and the self-coupling $\lambda$ are restricted to positive values to ensure 
a well behaved graviton and a stable minimum, respectively. Performing the Weyl transformation\,\footnote{Although
used extensively in the literature, we refrain from calling a pointwise rescaling of the metric ``conformal transformation.'' For
more details on the differences between Weyl and conformal invariance, see~\cite{Karananas:2015ioa}. }
$g_{\mu\nu}\rightarrow M_P^2/(\xi_h h^2) g_{\mu\nu}$ and defining the dimensionless
variables $Z^{-1}=\xi_h h^2/M_P^2$ and $\Phi=\chi/M_P$, the induced gravity 
Lagrangian~\eqref{Hig-Inf-Ind2} can be written in the so-called Einstein frame as
\be{act_ind}
\begin{aligned}
\frac{\mathscr L}{\sqrt{g}}=\frac{M_P^2}{2}R 
-\frac{M_P^2}{2}&\Big[ K_{ZZ}(Z)(\partial Z)^2+K_{\Phi\Phi}(Z)(\partial\Phi)^2 \Big]-\frac{\lambda M_P^4}{4\xi_h^2}\ ,
\end{aligned}
\ee
with
\be{functKind}
\begin{aligned}
K_{ZZ}(Z)=-\frac{1}{4\kappa_c} \frac{1}{Z^2}\,,\hspace{15mm}
K_{\Phi\Phi}(Z)=Z \ ,
\end{aligned}
\ee
and
\be{kap_c}
\kappa_c \equiv-\frac{\xi_h}{1+6\xi_h}\ .
\ee

An interesting observation is that the coefficient functions $K_{ZZ}(Z)$ and $K_{\Phi\Phi}(Z)$ are not actually independent. This 
property allows to rewrite~\eqref{act_ind}  in terms of $K_{\Phi\Phi}(Z)$
only\, \footnote{Note that we have dropped the argument $Z$ in $K_{\Phi\Phi}$ to stress that $K_{\Phi\Phi}$ 
\textit{itself} is the relevant variable for inflation.}
\be{act_ind2}
\frac{\mathscr L}{\sqrt{g}}=\frac{M_P^2}{2}R 
-\frac{M_P^2}{2}\Bigg[-\frac{\text{sign}({\kappa_c})(\partial K_{\Phi\Phi})^2}{4\,\vert \kappa_c\vert\, 
K^2_{\Phi\Phi}} +K_{\Phi\Phi}(\partial \Phi)^2 \Bigg]-\frac{\lambda M_P^4}{4\xi_h^2}\ .
\ee
This way of writing the Lagrangian is particularly enlightening, for it provides a physical 
interpretation for the \emph{constant} $\kappa_c$\,: it is the Gaussian curvature (in units of $M_P$) 
of the manifold spanned by the coordinates $K_{\Phi\Phi}$ and $\Phi$, as can be easily checked by an 
explicit computation. In the new language, the requirement of healthy kinetic sector for the induced 
gravity scenario translates into $K_{\Phi\Phi}>0$ and $\kappa_c<0$. The second condition implies 
that the two-dimensional field manifold is hyperbolic. 

Note that it is possible to make the $K_{\Phi\Phi}$  kinetic term canonical 
by performing a field redefinition
\begin{equation}
\tilde Z=-\frac{M_P}{2\sqrt{\vert \kappa_c\vert}} \int \frac{d K_{\Phi\Phi}}{ K_{\Phi\Phi}}
=-\frac{M_P}{2\sqrt{\vert \kappa_c\vert}}\log K_{\Phi\Phi} \hspace{5mm}\rightarrow\hspace{5mm} 
K_{\Phi\Phi}=e^{-2\sqrt{\vert\kappa_c\vert}\frac{\tilde Z}{M_P}}\ .
\end{equation}
The minus sign in this expression ensures that $\tilde Z$ goes to zero at $K_{\Phi\Phi}=0$. In terms of the 
canonically normalized variable $\tilde Z$, the theory~\eqref{act_ind2} reads
\be{}
\frac{\mathscr L}{\sqrt{g}}=\frac{M_P^2}{2}R 
-\frac{1}{2}\Bigg[(\partial \tilde Z)^2+
e^{-2\sqrt{\vert\kappa_c\vert}\frac{\tilde Z}{M_P}} (\partial \chi)^2 \Bigg]-\frac{\lambda M_P^4}{4\xi_h^2} \ .
\ee

\section{The minimal scale-invariant model}\label{sec:HDmodel}
The particular choice of functions in the induced gravity scenario ($f(h)\propto \sqrt{U(h)}$) 
gives rise to a constant potential which does not allow for a graceful inflationary exit. This problem can be easily overcome by introducing interactions for the $\chi$ field. Within a 
scale-invariant framework, the simplest possibility is to consider 
\be{HD-action-2}
\frac{\mathscr L}{\sqrt{g}}=\frac{\jx \x^2+\jh h^2}{2}R-\frac{1}{2}(\partial h)^2
-\frac{1}{2}(\partial \chi)^2-\frac{\l}{4}\left(h^2-\a\x^2 \right)^2-\b \x^4 \ ,
\ee
with $\alpha$ and $\beta$ constants. In what follows, we will restrict ourselves to positive or zero values of the 
non-minimal couplings $\xi_h$ and $\xi_\chi$. As in the previous section, this choice ensures that the graviton propagator is properly normalized for all field values. 

 The inflationary dynamic of this extended theory is more easily understood in a different 
set of variables 
\be{pol-var}
\Phi^2=\xi_h h^2+\xi_\chi\chi^2\ ,~~~\text{and}~~~
Z^{-1}=1+\frac{\xi_h h^2}{\xi_\chi\chi^2} \ ,
\ee
which are positive-definite for $\xi_h,\xi_\chi\geq0$. In terms of the fields $(\Phi,Z)$, the model~\eqref{HD-action-2} becomes
\be{HD-action-3}
\begin{aligned}
\frac{\mathscr L}{\sqrt{g}} = \frac{ \Phi^2 R}{2}
-\frac{ \Phi^2}{2}&\Big[ G_{ZZ}(Z)(\partial Z)^2 +
2   G_{Z\Phi}(Z) (\partial Z)\left(\Phi^{-1}\partial \Phi \right)
\\
&+G_{\Phi\Phi}(Z)\left(\Phi^{-1}\partial\Phi\right)^2\Big]- 
\Phi^4 v(Z)\ ,
\end{aligned}
\ee
with
\be{functG}
\begin{aligned}
G_{ZZ}(Z)&=\frac{1-\xi_\chi Z G'_{\Phi\Phi}(Z)}{4\xi_\chi Z(1-Z)}\ ,\hspace{17mm} 
G_{Z\Phi}(Z)= \frac{1}{2}G'_{\Phi\Phi}(Z)\ ,\hspace{5mm} \\
G_{\Phi\Phi}(Z)&=\frac{1}{\xi_h}+ \frac{\xi_h-\xi_\chi}{\xi_h\xi_\chi} Z\ ,\hspace{20mm}
v(Z)=\frac{\lambda}{4\jh^2}\left(1-Z-\alpha\,\frac{\xi_h}{\xi_\chi} Z\right)^2+\frac{\beta}{\xi_\chi^2}Z^2 \ ,
\end{aligned}
\ee
and the primes denoting derivatives with respect to $Z$. For $\Phi^2 \neq 0$, the Lagrangian~\eqref{HD-action-3} can be transformed to the Einstein frame by rescaling the metric as $g_{\mu\nu}\rightarrow M_P^2/\Phi^2 g_{\mu\nu}$. Doing this, we obtain
\be{act_max_symHD}
\begin{aligned}
\frac{\mathscr L}{\sqrt{g}}=\frac{M_P^2}{2}R 
-\frac{M_P^2}{2}&\Big[ K_{ZZ}(Z)(\partial Z)^2+
2K_{Z\Phi}(Z)(\partial Z)(\partial\log\Phi/M_P)\\
&+K_{\Phi\Phi}(Z)(\partial\log\Phi/M_P)^2 \Big]-V(Z) \ ,
\end{aligned}
\ee
with 
\be{functK}
\begin{aligned}
K_{ZZ}(Z)&=\frac{1-\xi_\chi Z K'_{\Phi\Phi}(Z))}{4\xi_\chi Z(1-Z)}\ ,~~~~\hspace{15mm}
K_{Z\Phi}(Z)=\frac{1}{2}K'_{\Phi\Phi}(Z)\ ,\\ 
K_{\Phi\Phi}(Z)&=6+G_{\Phi\Phi}(Z)\ ,~~~~~~ \hspace{23mm}
V(Z)=M_P^4\, v(Z) \ .
\end{aligned}
\ee
Note that the two-field model presented in~\eqref{act_max_symHD} displays some important differences with respect to the 
induced gravity scenario considered in Sec.~\ref{sec:IG}. The role of the gravitational interactions of the $\chi$ field 
is twofold. On the one hand, they induce a running on the inflationary potential, which now deviates  
from a constant even if $\alpha=\beta=0$. On the other hand, they give rise to a nontrivial kinetic 
mixing between the fields. In order to get rid of this mixing,
it suffices to consider a shift of the $\Phi$ field by
\be{dil_shift}
\log\frac{\Phi}{M_P}\rightarrow \log\frac{\Phi}{M_P}- \varphi(Z)\ ,~~~~\text{with}~~~~
\varphi'(Z)=\frac{K_{Z\Phi}(Z)}{K_{\Phi\Phi}(Z)} \ .
\ee
After this shift, Eq.~\eqref{act_max_symHD} becomes
\be{act_max_sym_HD2}
\frac{\mathscr L}{\sqrt{g}}=\frac{M_P^2}{2}R 
-\frac{M_P^2}{2}\Bigg[K(Z)(\partial Z)^2+K_{\Phi\Phi}(Z)(\partial \log\Phi/M_P)^2 \Bigg]-V(Z) \ ,
\ee
with
\be{F}
K(Z)= \frac{K_{ZZ}(Z)K_{\Phi\Phi}(Z)-K_{Z\Phi}^2(Z)}{K_{\Phi\Phi}(Z)}\ .
\ee 
The resulting Lagrangian, albeit diagonal, still contains two functions. 
Note however, that they are not really independent, since scale invariance forces 
them to depend on the dimensionless variable $Z$ only.  
Using the explicit expressions~\eqref{functK}, the coefficient $K(Z)$ in~\eqref{F} becomes
\be{FK0}
K(Z)=
\frac{1}{4 Z(Z-\zeta)}\left[6-\frac{1+6\kappa_0}{\kappa_0}\frac{1}{1-Z}\right] \ ,
\ee
with 
\be{kap_0}
\kappa_0 \equiv\kappa_c\left(1-\frac{\xi_\chi}{\xi_h}\right)\,, \hspace{10mm}
\zeta\equiv \frac{\kappa_0-\kappa_c}{\kappa_0(1+6\kappa_c)}\ ,
\ee
and $\kappa_c$ the induced gravity curvature defined in Sec.~\ref{sec:IG}. When written this way, 
it becomes clear that the two-field model under consideration shares some properties with 
the attractor models discussed in~\cite{Galante:2014ifa} and studied in detail in a number of papers, see for example~\cite{Kallosh:2013daa,Broy:2015qna,Kallosh:2016gqp} and references therein. Let us note that~\eqref{FK0} displays three poles: an 
inflationary pole at $Z=0$, a ``Minkowski'' pole at $Z=1$ and a pole at $Z=\zeta$. The condition $\xi_h>\xi_\chi>0$
guarantees that both $K(Z)$ and $K_{\Phi\Phi}(Z)$ are positive-definite in the interval 
$0< Z< 1$,\footnote{Note that
\begin{equation}
\lim_{Z\rightarrow0^+} K(Z)=1/\text{sign}(\xi_\chi)\ ,\hspace{10mm}
\lim_{Z\rightarrow1^-} K(Z)=1/\text{sign}(\xi_h) \ .
\end{equation}}
and makes unreachable the pole at $Z=\zeta$.\footnote{Let us mention  
that $-1/6\leq\kappa_c<\kappa_0<0$ for $\xi_h>\xi_\chi>0$. The pole appears at negative 
$Z$ while $0<Z<1$ in this case.} 
For field values 
relevant for inflation we have $Z\ll 1$, so Eq.~\eqref{FK0} can be approximated by
\be{FK0app}
K(Z)\approx -
\frac{1}{4 \kappa_0 Z(Z-\zeta)}  \ .
\ee
Using the expressions~\eqref{functG} and~\eqref{functK}, we find that 
$Z=\zeta\cdot\left(\kappa_c K_{\Phi\Phi}(Z)+1\right)$. This allows us to recast 
Eq.~\eqref{FK0app} in terms of $K_{\Phi\Phi}(Z)$ only
\be{FK0d}
K(Z)=-\frac{(\kappa_c K^{'}_{\Phi\Phi}(Z))^2}{4 \kappa_0(\kappa_c K_{\Phi\Phi}(Z))
(\kappa_c K_{\Phi\Phi}(Z)+1)}\ .
\ee
The form of \eqref{FK0d} is clearly reminiscent of that appearing  in the induced gravity scenario, 
cf. Eq.~\eqref{act_ind2}. The analogy can be made even more explicit once we define
\be{}
\tilde K_{\Phi\Phi}(Z)\equiv  \left\vert \frac{\kappa_c}{\kappa_0}\right\vert  K_{\Phi\Phi}(Z)
\ , \hspace{7mm} 
\tilde \Phi\equiv\sqrt{ \left\vert \frac{\kappa_0}{\kappa_c}\right\vert  }\log\frac{\Phi}{M_P}\,,
\ee
in terms of which, Eq.~\eqref{act_max_sym_HD2} takes the form\,\footnote{As we did in Sec.~\ref{sec:IG}, we drop the argument of $K_{\Phi\Phi}$, which now becomes the new field variable.}
\begin{eqnarray}\label{HDfinalact0}
\frac{\mathscr L}{\sqrt{g}}&=&\frac{M_P^2}{2}R -
\frac{M_P^2}{2}\Bigg[\frac{(\partial \tilde K_{\Phi\Phi})^{2}}{4\, 
\,\tilde K_{\Phi\Phi}
(\vert\kappa_0\vert \tilde K_{\Phi\Phi}-1)}+\tilde 
K_{\Phi\Phi}(\partial \tilde \Phi)^2 \Bigg] -V(K_{\Phi\Phi})\ ,
\end{eqnarray}
with
\be{VKHD}
V(K_{\Phi\Phi})=V_0\left(1 -\sigma \vert\kappa_0\vert\, \tilde K_{\Phi\Phi}\right)^2 +
\frac{\beta}{\vert \kappa_0 \vert^2} \left(1-\vert \kappa_0 \vert \tilde K_{\Phi\Phi}\right)^2 \ ,
\ee
and  
\be{}
V_0\equiv\frac{\lambda \, a^2 M_P^4}{4}
\ ,\hspace{8mm}  a\equiv -\frac{1-6 \vert\kappa_0\vert}{\vert\kappa_0\vert}-\frac{\alpha}{\vert\kappa_0\vert}\ ,
\hspace{8mm} \sigma \vert\kappa_0\vert \equiv -\frac{1}{a}
\left(\frac{\vert \kappa_c\vert-\vert \kappa_0\vert}{\vert \kappa_c\vert }+\alpha\right)\ ,
\ee
The interpretation of the quantities defined in our derivation is now straightforward:
\begin{enumerate}
 \item $\kappa_0$ can be interpreted as the Gaussian curvature of 
the field-derivative manifold~\eqref{HDfinalact0}, which becomes maximally symmetric around the inflationary pole $Z=0$. It should be stressed that the curvature of the manifold is in general not constant and reads
\be{kHD0}
\kappa=\kappa_0\left[1-2\left(1-6\vert\kappa_0\vert\right)Z\right]\ .
\ee
\item $\sigma \vert\kappa_0\vert $ contains two pieces associated with  the gravitational and potential 
interactions between the two scalar fields, respectively. For $\alpha=0$, $\sigma \vert\kappa_0\vert$ measures (up to some normalization) 
the difference between  the Gaussian curvature of  the field-derivative manifold \eqref{HDfinalact0} and
the induced gravity curvature $\kappa_c$.
\end{enumerate}
The kinetic terms in Eq.~\eqref{HDfinalact0} can be made canonical by considering an additional field redefinition
\be{KppfinalHD}
\tilde Z=\int dZ\sqrt{K(Z)}  \hspace{10mm} \longrightarrow  \hspace{10mm}\tilde K_{\Phi\Phi}=
\frac{1}{\vert \kappa_0\vert}\cosh^2 \frac{\sqrt{\vert\kappa_0\vert}\tilde Z}{M_P}\,.
\ee
Doing this, we get 
\begin{eqnarray}\label{action_HD}
\frac{\mathscr L}{\sqrt{g}}&=&\frac{M_P^2}{2}R 
-\frac{M_P^2}{2}\Bigg[(\partial \tilde Z)^2 +\frac{1}{\vert\kappa_0\vert}\cosh^2\frac{\sqrt{\vert\kappa_0\vert}\tilde Z}{M_P}
(\partial \tilde\Phi)^2 \Bigg] - V(\tilde Z)\ , 
\end{eqnarray}
where the potential reads
\be{potHD}
V(Z)=V_0\left(1 -\sigma \cosh^2 \frac{\sqrt{\vert\kappa_0\vert}\tilde Z}{M_P}\right)^2+
\frac{\beta}{\vert \kappa_0 \vert^2} \left(1-\cosh^2\frac{\sqrt{\vert\kappa_0\vert}\tilde Z}{M_P}\right)^2 \ .
\ee

Note that 
for $\alpha,\beta \ll 1$, one recovers the Higgs-dilaton Lagrangian found 
in Ref.~\cite{GarciaBellido:2011de} 
(see also~\cite{Shaposhnikov:2008xb,Shaposhnikov:2008xi,Bezrukov:2012hx,GarciaBellido:2012zu,Rubio:2014wta,
Trashorras:2016azl}). The Jordan frame formulation of this model has been recently revisited in 
Ref.~\cite{Ferreira:2016vsc}.

\section{General scale invariant models: the maximally symmetric case}\label{sec:gencase}

Here, we generalize the results of Sec.~\ref{sec:HDmodel} to general scale-invariant models of 
inflation involving two scalar degrees of freedom. The most general Lagrangian density containing 
terms which are at most quadratic in the derivatives is given by
\be{TdiffL0}
\begin{aligned}
\frac{\mathscr L}{\sqrt{ g}}=
\frac{\Phi^2f(Z)}{2} R-\frac{\Phi^2}{2}&\Big[G_{ZZ}(Z)\left(\partial Z\right)^2+2G_{Z\Phi}(Z)
\left(\partial Z\right)\left(\Phi^{-1}\partial\Phi \right)\\
&+G_{\Phi\Phi}(Z)\left(\Phi^{-1}\partial \Phi \right)^2
\Big]-\Phi^4v(Z) \ .
\end{aligned}
\ee
The functions $f(Z),G_{ZZ}(Z),G_{\Phi\Phi}(Z),G_{Z\Phi}(Z)$ and $v(Z)$ in this expression are arbitrary 
functions of $Z$ only (not necessarily polynomials). They can either be introduced \textit{ad hoc}, or emerge naturally 
in the context of modified gravitational theories. A particular example of the second possibility appears in theories which are invariant under 
transverse diffeomorphisms (TDiff), a restricted group of general coordinate transformations preserving the four-volume, 
see for instance~\cite{Blas:2011ac,Karananas:2016grc}.\footnote{TDiff theories depend on arbitrary functions of the 
metric determinant and generically contain an additional scalar mode in the gravitational sector. In order to study the dynamics 
of these models it is useful to formulate them in a diffeomorphism-invariant language by introducing a St\"uckelberg field. When 
this is done, the additional degree of freedom appears explicitly in the Lagrangian.}
 
For $\Phi^2f(Z) \neq 0$, we can get rid of the nonlinearities in the gravitational sector of~\eqref{TdiffL0}, by Weyl-transforming 
the metric $g_{\mu\nu}\rightarrow M_P^2/\Phi^2f(Z) g_{\mu\nu}$.
The resulting Lagrangian reads
\be{act_max_sym}
\begin{aligned}
\frac{\mathscr L}{\sqrt{g}}=\frac{M_P^2}{2}R 
-\frac{M_P^2}{2}&\Big[ K_{ZZ}(Z)(\partial Z)^2+
2K_{Z\Phi}(Z)(\partial Z)(\partial\log\Phi/M_P)\\
&+K_{\Phi\Phi}(Z)(\partial\log\Phi/M_P)^2 \Big]-V(Z) \ ,
\end{aligned}
\ee
with
\begin{eqnarray}\label{Ks0}
K_{ZZ}(Z)&=&\dfrac{G_{ZZ}(Z)}{f(Z)}+\dfrac{3}{2}\left(\dfrac{
f'(Z)}{f(Z)}\right)^2\,,\hspace{10mm}
K_{Z\Phi}(Z)=\dfrac{G_{Z\Phi}(Z)}{f(Z)}+3\,\dfrac{f'(Z)}{f(Z)}\ , \\
K_{\Phi\Phi}(Z)&=&6+\dfrac{G_{\Phi\Phi}(Z)}{f(Z)}\ , \hspace{31mm}
V(Z)=\dfrac{M_P^4  \,v(Z)}{f^2(Z)}\ , 
\end{eqnarray}
and the primes denoting derivative with respect to $Z$. Although 
$K_{ZZ}(Z)$, $K_{Z\Phi}(Z)$ and $K_{\Phi\Phi}(Z)$ are in principle
arbitrary, some physical requirement, such as the absence 
of ghosts in the spectrum, can significantly reduce the number of admissible functions. Diagonalizing the 
kinetic terms in~\eqref{act_max_sym} by shifting the dilaton field $\Phi$ as in \eqref{dil_shift}, we get
\be{act_max_sym_2}
\frac{\mathscr L}{\sqrt{g}}=\frac{M_P^2}{2}R 
-\frac{M_P^2}{2}\Bigg[K(Z)(\partial Z)^2+K_{\Phi\Phi}(Z)(\partial \log\Phi/M_P)^2 \Bigg]-V(Z) \ ,
\ee
with 
\be{F2}
K(Z)=
\frac{K_{ZZ}(Z)K_{\Phi\Phi}(Z)-K_{Z\Phi}^2(Z)}{K_{\Phi\Phi}(Z)}\ .
\ee 
Once again, the absence of ghosts translates into a condition on the functions $K_{\Phi\Phi}(Z)$ and $K(Z)$, which are 
required to be positive-definite
\be{semipos}
K(Z)>0\ ,\hspace{10mm}K_{\Phi\Phi}(Z)>0\ .
\ee
The kinetic sector of Eq.~\eqref{act_max_sym_2} constitutes a nonlinear sigma model. The associated 
Gaussian curvature in Planck units is given by
\be{gauss}
\kappa(Z)
=\frac{K_{\Phi\Phi}'(Z)F'(Z)-2F(Z) K''_{\Phi\Phi}(Z)}{4F^2(Z)} \ ,
\ee
where, in order to keep the notation short, we have defined $F(Z)\equiv K(Z)K_{\Phi\Phi}(Z)$.
For inflationary models in which $\kappa(Z)$ is approximately 
constant during inflation, Eq.~\eqref{gauss} can be easily integrated to obtain 
\be{req_gho_free_2} 
K(Z)=-\frac{ K^{'2}_{\Phi\Phi}(Z)}{4 \, K_{\Phi\Phi}(Z)
( \kappa   K_{\Phi\Phi}(Z)+c)}\,,
\ee
with $c$ an integration constant. The associated Lagrangian density reads\footnote{We assumed 
that the potential is an analytic function of $Z$, such that it can be expressed in term of $K_{\Phi\Phi}$ as well.}
\begin{equation}\label{MSattractor}
\frac{\mathscr L}{\sqrt{g}}=\frac{M_P^2}{2}R -
\frac{M_P^2}{2}\Bigg[-\frac{(\partial  K_{\Phi\Phi})^{2}}{4\,  K_{\Phi\Phi}
(\kappa K_{\Phi\Phi}+c)}+ 
K_{\Phi\Phi}(\partial  \Phi)^2 \Bigg] -V(K_{\Phi\Phi}) \,, 
\end{equation}
with arbitrary (but positive-definite) function $K_{\Phi\Phi}$.

The target manifold of this family of models is maximally symmetric for all values of $c$. Depending on whether $\kappa$ 
is positive or negative, the geometry of the field-derivative space corresponds to a sphere or a to a Gauss-Bolyai-Lobachevsky 
space. Different choices of $c$ can be associated with different models within class. Note that in order to ensure the absence of ghosts in 
the spectrum, the conditions $\kappa K_{\Phi\Phi}+c<0$ and $K_{\Phi\Phi}>0$ must be satisfied. The choices of parameters 
and field ranges fulfilling these requirements are summarized in Table~\ref{table1}.
The induced gravity model \eqref{act_ind2} and the two-field model~\eqref{HDfinalact0} are just two particular 
examples of the cases II and III.

It should be mentioned that contrary to what happens in single-field models, scale invariance does not seem to guarantee the emergence 
of an approximately shift symmetric potential in the Einstein frame. Asymptotically flat potentials as those appearing 
in the Starobinsky or Higgs inflation models are recovered only in the $c=0$ case. For $c\neq 0$, the inflationary region 
is limited to a compact field range. This can be seem explicitly by canonically normalizing the $K_{\Phi\Phi}$ kinetic term. 
Consider for instance the cases III and IV in Table \ref{table1}. Inserting  into~\eqref{MSattractor} the field redefinition 
\be{Kppfinal} 
 K_{\Phi\Phi}=\frac{c}{-\kappa}\cosh^2\left(\frac{\sqrt{-\kappa}\tilde Z}{M_P}\right)\ ,
\ee
with the restriction $\textrm{sign}(c)=\textrm{sign}(-\kappa)$, we get
\begin{equation}\label{MSattractor1} 
\frac{\mathscr L}{\sqrt{g}}=\frac{M_P^2}{2}R 
-\frac{M_P^2}{2}\Bigg[(\partial \tilde Z)^2  +\frac{c}{-\kappa}\cosh^2\left(\frac{\sqrt{-\kappa}\tilde Z}{M_P}\right)
(\partial \tilde\Phi)^2 \Bigg] - V\left[\frac{c}{-\kappa}\cosh^2\left(\frac{\sqrt{-\kappa}\tilde Z}{M_P}\right)\right]\ .
\end{equation}
The functional form of the potential depends also on the sign of the curvature. For $\kappa<0$, the potential is 
constructed out of hyperbolic functions, while for $\kappa>0$ one rather gets natural-like inflation potentials~\cite{Freese:1990rb}. 

\begin{table}
\begin{center}
\begin{tabular}{l cc c cc c cc l}
\toprule
Case I &&& $\kappa=0$ &&& $c<0$ &&& $K_{\Phi\Phi}>0$ \\
Case II &&&$\kappa <0$ &&& $c\leq 0$ &&& $K_{\Phi\Phi}>0$~\footnote{Note that the restriction $\kappa K_{\Phi\Phi}+c<0$ is
satisfied for $K_{\Phi\Phi}>-\vert c \vert/\vert \kappa\vert $. However, as already stated, negative values of $K_{\Phi\Phi}$ must 
be avoided in order to have a well-normalized dilaton.}  \\
Case III &&&$\kappa <0$ &&& $c>0$  &&& $K_{\Phi\Phi}>\frac{c}{-\kappa}$ \\
Case IV &&& $\kappa >0$ &&& $c<0$&&& $\frac{-c}{\kappa}>K_{\Phi\Phi}>0$\\ 
\bottomrule 
\end{tabular}
\end{center}
\caption{Restrictions of the kinetic sector of maximally symmetric two-field models of inflation 
ensuring the absence of ghosts.}\label{table1}
\end{table}

The inflationary observables of the maximally symmetric model~\eqref{MSattractor} are 
determined by the pole structure of the $K_{\Phi\Phi}$  kinetic term. For concreteness, we will concentrate 
on the case III, which is the one appearing in the simplest modification of the induced gravity scenario. The analysis of the other cases presented in Table~\ref{table1} goes along the same lines.

For $\vert c\vert \rightarrow 0$, the stability of the $K_{\Phi\Phi}$ kinetic term forces 
$K_{\Phi\Phi}\rightarrow \vert c/ \kappa\vert \rightarrow 0$ (cf. Table~\ref{table1}) and the $K_{\Phi\Phi}$ pole becomes essentially
quadratic. The spectral tilt $n_s$ and tensor-to-scalar ratio $r$ coincide in this limit with those 
in Refs.~\cite{Kallosh:2013yoa,Galante:2014ifa}. As in that case, the details of the model (choice of functions, 
shape of the potential, etc \dots) do not affect the inflationary predictions at the lowest order in the (inverse)
number of e-folds $N$ 
\be{}
n_s\approx 1-\frac{2}{N} \ , \hspace{15mm}
r\approx \frac{2}{\vert \kappa \vert N^2}  \ .
\ee

For $|c|\neq 0$, the inflationary pole at $K_{\Phi\Phi}=0$ is no longer reachable. Around the pole at $K_{\Phi\Phi}\approx c/\vert\kappa\vert$, the Lagrangian density~\eqref{MSattractor} 
can be approximated by
\begin{equation}\label{MSattractorapprox}
\frac{\mathscr L}{\sqrt{g}}\approx\frac{M_P^2}{2}R -
\frac{M_P^2}{2}\left(\frac{1}{4\vert c\vert\left(K_{\Phi\Phi}-\frac{\vert c\vert}{\vert\kappa\vert}\right)}+\ldots \right)
(\partial  K_{\Phi\Phi})^{2}-V_0\left[1-\sigma_0 \left(K_{\Phi\Phi}-\frac{\vert c \vert}{\vert\kappa\vert}\right)
+\ldots\right] \ ,
\end{equation}
with the ellipses denoting higher order terms and $V_0$ an overall 
coefficient to be fixed by observations. The normalization constant  $\sigma_0$ in the potential can be set to 
one without loss of generality. Note indeed that the particular structure of the 
$K_{\Phi\Phi}$ kinetic term in~\eqref{MSattractor} and~\eqref{MSattractorapprox}  allows to absorb $\sigma_0$ into the 
definition of $c$ by performing a scaling  $K_{\Phi\Phi}\rightarrow K_{\Phi\Phi}/\sigma_0$,  $\vert c \vert
\rightarrow \vert c\vert/\sigma_0$. Thus, there are only three independent parameters, namely $V_0$, $\kappa$ and $c$.

The kinetic sector of~\eqref{MSattractorapprox} contains a linear pole. As shown in Ref.~\cite{Terada:2016nqg}, the 
spectral tilt and the tensor to scalar ratio in this case asymptote the values
$n_s\rightarrow -\infty$ and $r\rightarrow 0$ in the large $\vert c\vert$ limit. 
These results generalize the predictions of the simplest Higgs-Dilaton model~\eqref{action_HD}, to a general class of 
theories in which the defining functions in~\eqref{TdiffL0}  give rise to an Einstein-frame target manifold 
with approximately constant curvature during inflation. The multifield cosmological attractors considered in 
Ref.~\cite{Kallosh:2013daa} are also a particular case within this category, with $\vert\kappa\vert=1/6$.\footnote{Note that contrary to the cases considered in 
that work, the potential in our case is restricted to depend on a single field due to the presence of scale invariance.}

\section{Conclusions}
\label{sec:concl}

The purpose of this paper was to investigate how the geometrical properties of the 
target manifold affects the inflationary predictions of two-field scalar-tensor theories invariant under dilatations. 

To set the stage, we considered an induced gravity model with an additional non-interacting 
scalar degree of freedom. When this theory is written in the Einstein frame, the kinetic sector turns out to be 
noncanonical. The coefficients of the kinetic terms are however related by a 
very specific constraint that is provided by the maximally symmetric geometry of the field-derivative space.

Although the induced gravity model does not allow for a graceful inflationary exit, it provides us with a useful 
insight to move to the simplest viable inflationary model. This scenario contains two scalar fields non-minimally
coupled to gravity and polynomial interactions. We showed that during inflation,
the Einstein-frame kinetic terms are subject to the very same constraint than the ones in 
the induced gravity model. The interesting point is that 
the constant curvature of the field-manifold propagates now all the way to the inflationary observables.

Finally, by abandoning the requirement of polynomial interactions, we considered the most general 
scale-invariant theory involving no more than two derivatives. Since the Lagrangian of the model
contains (\textit{a priori}) five independent functions, making a general
statement about the inflationary predictions seems hopeless at first sight. However, we showed that if the 
corresponding target manifold is maximally symmetric during inflation, the 
 dynamics turn out to be completely controlled by a single function: the coefficient of the dilaton 
kinetic term in the Einstein frame. The particular pole structure of the kinetic sector makes the 
inflationary predictions insensitive to the details of the theory in the large number of e-folds limit and universal in the
sense of~\cite{Galante:2014ifa} . 

From this new perspective, the predictions of the Higgs-dilaton model~\cite{Shaposhnikov:2008xb,GarciaBellido:2011de}
are much more generic than what could be initially expected. In particular, they are not attached to a particular 
choice of functions, but they can be rather attributed to a defining principle. That is, a target manifold with 
approximately constant curvature during inflation.

\section*{Acknowledgements}

We would like to thank Mikhail Shaposhnikov for numerous discussions.
The work of G.K.K. is supported by the Swiss National Science Foundation.  J.R. acknowledges support from the
DFG through the project TRR33 ``The Dark Universe.''

\bibliographystyle{utphys}
\bibliography{Weyl_Higgs_Inflation}{}

\end{document}